\documentclass{spie}
\usepackage[english]{babel}
\usepackage{url}
\usepackage[bookmarksnumbered=true,breaklinks=true,pdfborder=0 0 0]{hyperref}
\usepackage{graphicx}
\usepackage{paralist}
\usepackage{subfig}
\usepackage{mathabx}
\usepackage{xfrac}
\usepackage{siunitx}
\usepackage[super]{natbib}

\newcommand{\figureone}[3]{%
\begin{figure}
\begin{center}
\includegraphics[width=4.75in]{#1}
\skiplinehalf
\caption{#3}
\label{#2}
\end{center}
\end{figure}
}

\newcommand{\figuretwo}[3]{%
\begin{figure}[tb]
\begin{center}
\includegraphics[width=6.5in]{#1}
\skiplinehalf
\caption{#3}
\label{#2}
\end{center}
\end{figure}
}

\newcommand{\figurethree}[5]{%
\begin{figure}[tb]
\begin{center}
	\subfloat{\includegraphics[width=1.75in]{#1}}\quad
	\subfloat{\includegraphics[width=1.75in]{#2}}\quad
	\subfloat{\includegraphics[width=1.75in]{#3}}
\skiplinehalf
\caption{#5}
\label{#4}
\end{center}
\end{figure}
}

\newcommand\arcsec{\si{\arcsecond}}
\newcommand\CaII{\mbox{Ca\,\textsc{ii}}}
\newcommand\HeI{\mbox{He\,\textsc{i}}}
\newcommand\Halpha{\mbox{H\ensuremath{\alpha}}}
\newcommand\FeI{\mbox{Fe\,\textsc{i}}}
\newcommand\FeIX{\mbox{Fe\,\textsc{ix}}}
\newcommand\nm{\si{\nano\meter}}
\newcommand\mm{\si{\milli\meter}}
\newcommand\cm{\si{\centi\meter}}
\newcommand\um{\si{\micro\meter}}
\newcommand\minute{\si{\minute}}
\newcommand\meter{\si{\meter}}
\newcommand\rad{\si{\radian}}
\newcommand\s{\si{\second}}

\def\solphys{Sol.~Phys.}%
\def\apj{ApJ}%
\def\araa{ARA\&A}%
\def\pasj{PASJ}%
\def\aap{A\&A}%
\def\apjl{ApJ}%
\def\ao{Appl.~Opt.}%

\title{The chromosphere and prominence magnetometer} 

\author{Alfred G. de Wijn, Christian Bethge, Steven Tomczyk, and Scott McIntosh
\skiplinehalf
National Ctr. for Atmospheric Research, P.O. Box 3000, Boulder, CO, USA
}

\authorinfo{Further author information: E-mail: dwijn@ucar.edu, Telephone: 1 303 497 2171}

\begin{document} 
\maketitle 

\begin{abstract}
The Chromosphere and Prominence Magnetometer (ChroMag) is conceived with the goal of quantifying the intertwined dynamics and magnetism of the solar chromosphere and in prominences through imaging spectro-polarimetry of the full solar disk.
The picture of chromospheric magnetism and dynamics is rapidly developing, and a pressing need exists for breakthrough observations of chromospheric vector magnetic field measurements at the true lower boundary of the heliospheric system.
ChroMag will provide measurements that will enable scientists to study and better understand the energetics of the solar atmosphere, how prominences are formed, how energy is stored in the magnetic field structure of the atmosphere and how it is released during space weather events like flares and coronal mass ejections.
An integral part of the ChroMag program is a commitment to develop and provide community access to the ``inversion'' tools necessary for the difficult interpretation of the measurements and derive the magneto-hydrodynamic parameters of the plasma.
Measurements of an instrument like ChroMag provide critical physical context for the Solar Dynamics Observatory (SDO) and Interface Region Imaging Spectrograph (IRIS) as well as ground-based observatories such as the future Advanced Technology Solar Telescope (ATST).
\end{abstract}


\section{INTRODUCTION}
\label{sec:intro}

The chromosphere is a deeply complex part of the solar atmosphere.
It is named after its red appearance just after the beginning and before the end of a total solar eclipse, caused by emission in the \Halpha\ line at $656.3~\nm$.
The same lines that show emission in the flash spectrum allow us to observe the chromosphere on the disk.
Early detailed observations
	\cite{Secchi1877}
already show the intricate structure of the chromosphere.
We now know that the magnetized chromosphere is permeated by ``spicules'' at the limb and their on-disk counterparts (``mottles'' and ``fibrils'')
	\cite{1968SoPh....3..367B}.
Filaments and prominences are ``clouds'' of chromospheric material supported by a complex magnetic field and embedded in the corona that may erupt and produce Coronal Mass Ejections (CMEs).

The chromosphere is of integral importance in the mass and energy balance of the outer solar atmosphere, with over 90\% of the non-radiative energy deposited there 
	\cite{1983ApJ...267..825W},
and requiring nearly one hundred times the mass and energy flux of the corona for sustenance
	\cite{1977ARA&A..15..363W}.
The chromosphere remains the most poorly understood region of the outer solar atmosphere.
The complex dynamics resulting from the interplay of magnetic field and convectively driven photospheric plasma is incredibly difficult to interpret unambiguously, yet critically important to our understanding of the mass and energy balance of the outer solar atmosphere, including such things as coronal heating, the solar wind, and energetic events collectively known as space weather.

The study of the chromosphere is the next frontier in solar physics.
The most fundamental unresolved issues in solar physics are tied to the chromosphere:
\begin{compactitem}
\item what are the processes that transport energy and mass in the outer solar atmosphere?
\item how do these processes accelerate the solar wind, and how do they heat the chromosphere and the corona?
\item how are prominences formed, how do they evolve, and how do they relate to CMEs?
\item how is energy stored, and how is it released during flares and CMEs? and
\item what is the nature of the changes in the magnetic structure of the outer solar atmosphere and heliosphere that accompany the 11-year solar cycle?
\end{compactitem}
The key to all these questions lies within the magnetic field, and finding the answers depends critically on measuring the chromospheric vector magnetic field and related dynamic processes over the full solar disk.

\figurethree{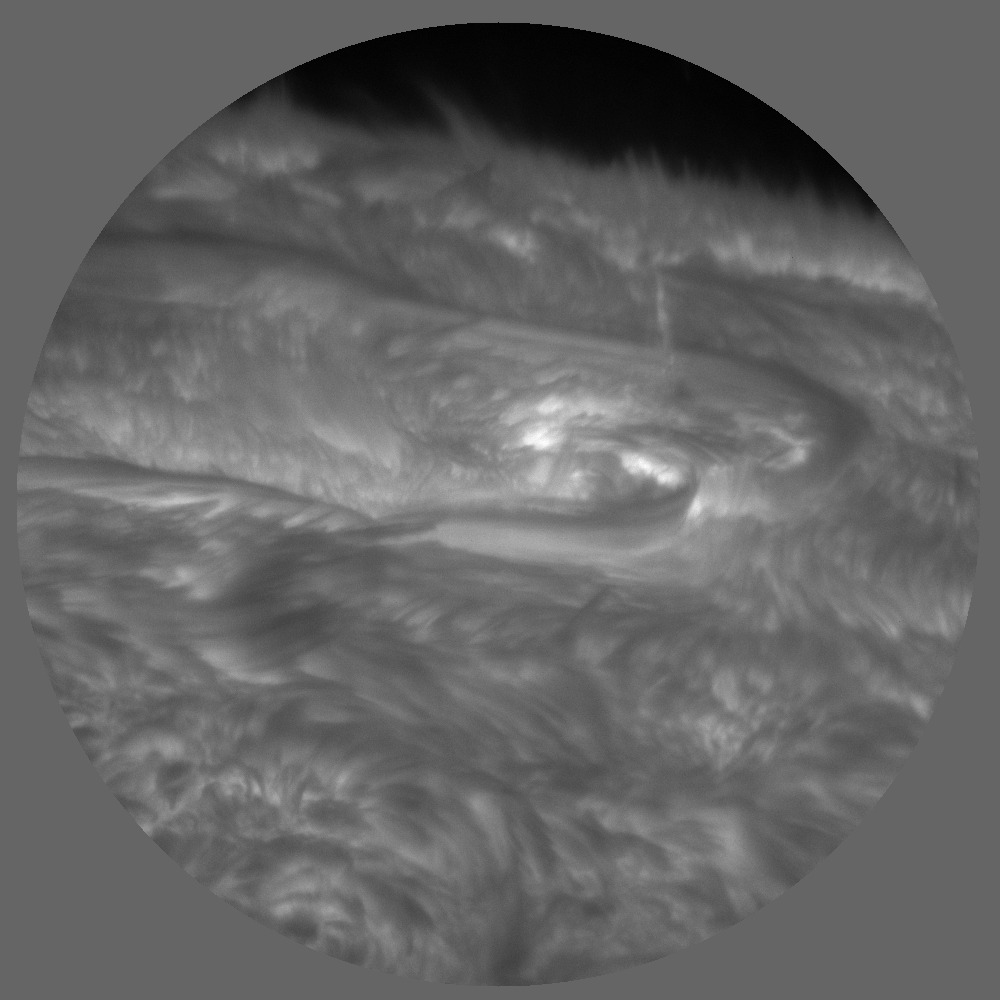}{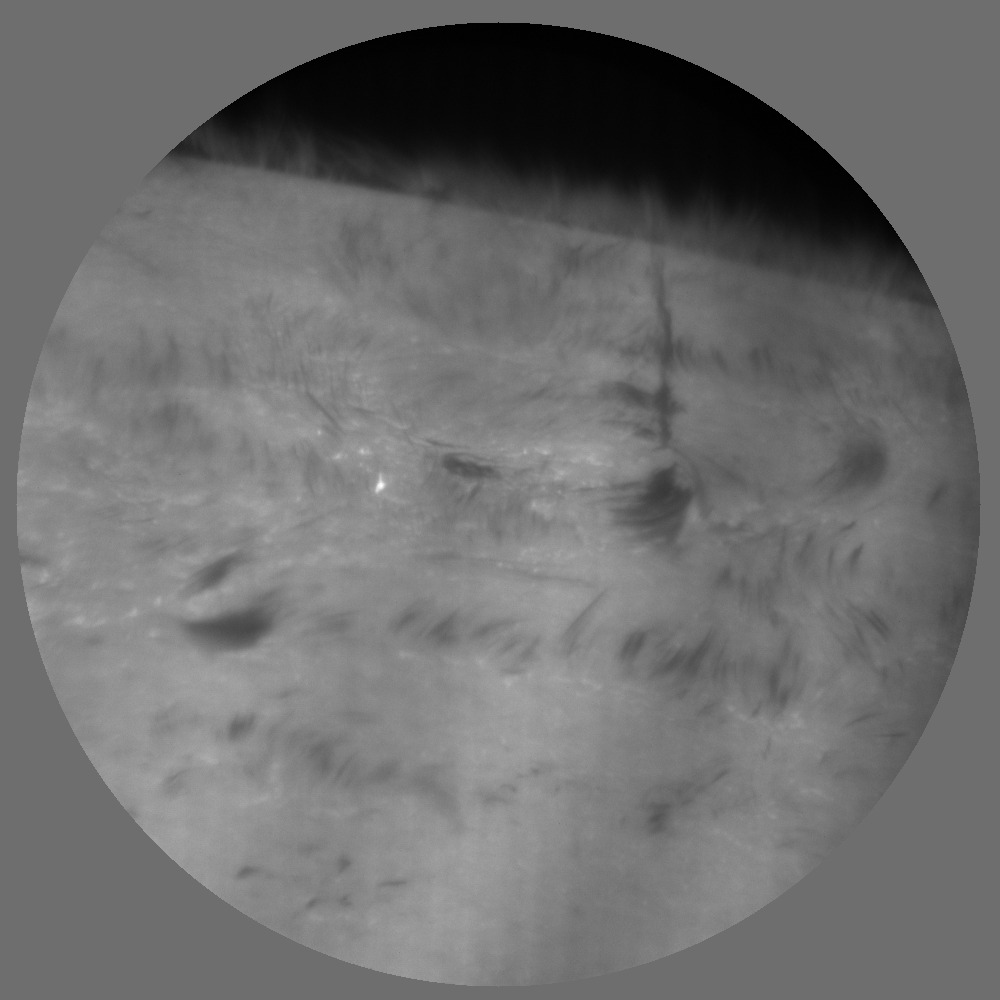}{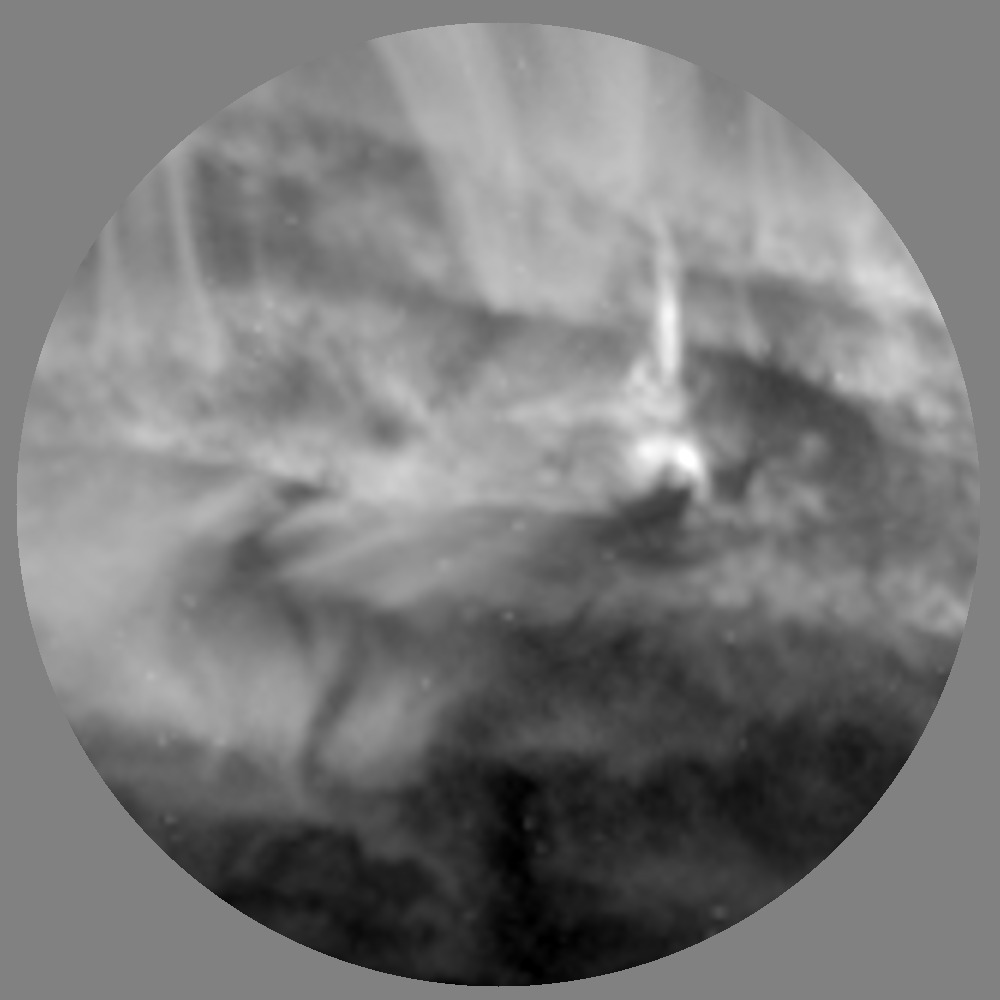}{fig:ibisaia}{IBIS images of \CaII\ $854.2~\nm$ line core (left) and $40$-$60~\mbox{km}/\mbox{s}$ in the blue wing (center) in comparison with \FeIX\ $17.1~\nm$ coronal emission observed by SDO Atmospheric Imaging Assembly (AIA) (right).
The circular field of view has a diameter of approximately $80\arcsec$.
Note the strong visual correspondence of the chromospheric filter images and the coronal emission.
The dark absorbing features in the center panel are known to be associated with a class of spicules that are thought to play a key role in the mass and energy balance of the corona and solar wind.}

Magnetism in the solar chromosphere is structured on all spatial scales.
At the very finest resolution, we find extremely dynamic features such as spicules, mottles, and fibrils, that can barely be resolved with the best telescopes in the world
	\cite{2007PASJ...59S.655D,2009ApJ...705..272R}.
An example of chromospheric fine-structure is shown in \autoref{fig:ibisaia}.
New telescopes, such as the 4-m Advanced Technology Solar Telescope (ATST), are being built specifically for high-resolution observations in order to study these kinds of features in detail.
High-resolution instruments like the Interferometric Bidimensional Spectrometer (IBIS)
	\cite{2006SoPh..236..415C}
on the Dunn Solar Telescope (DST) at Sacramento Peak and the CRisp Imaging SpectroPolarimeter (CRISP)
	\cite{2006A&A...447.1111S}
at the 1-m Swedish Solar Telescope on La Palma are already driving breakthroughs in our understanding of the complex environment of the chromosphere.
Those instruments however do not capture the larger scales.
EUV imagery from space missions like the Transition Region And Coronal Explorer (TRACE)
	\cite{1999SoPh..187..229H},
the Solar and Heliospheric Observatory (SOHO), and now the Solar Dynamics Observatory (SDO) show clearly the complex interconnectivity of the magnetic field over large distances between active regions and the quiet-Sun network.
It is clear from those observations that synoptic full-disk measurements of chromospheric magnetic field are needed to investigate the response of the atmosphere at the onset of space weather events like solar flares and CMEs.

Magnetic fields have been studied in the solar photosphere for many years by exploiting the Zeeman effect.
A degeneracy in atomic energy levels is removed in the presence of a magnetic field, causing a separation in the spectrum.
The Zeeman triplet consists of two shifted $\sigma$ components and one unshifted $\pi$ component, and the separation of these three components is a measure of the magnetic field strength.
In sunspots and active regions, the magnetic field is strong and the Zeeman splitting is comparable to, or even greater than, the line width for many well-known lines such as the \FeI\ lines around $630~\nm$.
For weaker fields, such as in the internetwork quiet Sun, the Zeeman splitting is small and the effect is only recognizable as a broadening of the line.
Fortunately, the Zeeman components are characterized by distinct polarization properties, and so the weaker magnetic fields can still be investigated by the use of polarimetric instruments.
While the first ``magnetographs'' measured only circular polarization, yielding only a diagnostic of line-of-sight component of the magnetic field, modern instruments are routinely used to infer the vector magnetic field in the photosphere through full-Stokes polarimetry, i.e., measurement of both circular and linear polarization properties across the spectral line.

In order to derive the magnetic field from the spectro-polarimetric data recorded by the instrument, it is processed with ``inversion'' codes that attempt to fit the line profile simultaneously in the four Stokes parameters with an atmosphere described by a dozen or more magnetohydrodynamic parameters.
Both the Hinode and the SDO missions have magnetograph instruments on board that employ this method for routine measurements of photospheric magnetic field.
The Spectro-Polarimeter (SP)\cite{2008SoPh..249..167T} instrument on Hinode is a traditional spectrograph polarimeter that routinely produces measurements with unprecedented spatial resolution and polarimetric accuracy using the \FeI\ $630.15$ and $630.25~\nm$ line pair.
The Helioseismic and Magnetic Imager (HMI) on SDO employs full-disk imaging spectro-polarimetry in the \FeI\ $617.3~\nm$ line to produce field measurements at a regular cadence of $45~\s$.
Despite the urgent need for chromospheric vector magnetic field measurements, neither of these instruments nor the upcoming Interface Region Imaging Spectrograph (IRIS) have the capability to make those measurements.

The magnetic field in the chromosphere is generally much weaker than the photospheric field.
As a result, the two polarized lobes of the Zeeman effect largely cancel.
In addition, lines that sample the chromosphere are never formed in local thermal equilibrium (LTE), and the departure from LTE typically includes atomic polarization.
A full treatment of the magnetic effects on line polarization is required, including level-crossing interferences and the Hanle effect.
While the Zeeman effect is predominantly sensitive to longitudinal magnetic field, the Hanle effect is sensitive to transverse magnetic field.
Inversion codes that exploit the complementarity of the Zeeman and Hanle diagnostics, such as the HAZEL code developed at the Instituto de Astrof\'\i{}sca de Canarias in Spain are becoming available now.
Magnetic field has been successfully measured in regions where the plasma is relatively cool, such as prominences and filaments, or where the field is relatively strong, such as active regions
	\cite{2003ApJ...598L..67C,2009A&A...501.1113K}.

The Vector Spectromagnetograph (VSM) on the ground-based Synoptic Optical Long-term Investigations of the Sun (SOLIS)
	\cite{2003SPIE.4853..194K}
telescope makes regular observations in the chromospheric \CaII\ $854.2~\nm$ and \HeI\ $1083.0~\nm$ lines.
However, the VSM only measures intensity in both lines and circular polarization in \CaII\ $854.2~\nm$, yielding just a line-of-sight diagnostic.
Furthermore, since it is a slit-scanning spectrograph, the cadence of the observations is too slow to study the dynamics of the chromosphere and transient events like flares.

Summarizing, there is a considerable need for regular observations of the chromospheric vector magnetic field.
Such measurements complement the capabilities of the Hinode, SDO, and IRIS missions and improve our ability to interpret the observations from those missions.
Hence, the heliophysics community eagerly awaits the development of an instrument that will provide the community with synoptic global observations of the chromospheric vector field.

\section{INSTRUMENT DESCRIPTION AND REQUIREMENTS}
\label{sec:instrument}

\begin{table}
	\caption{Instrument Requirements and Goals.  The prototype instrument design specifications are shown in the rightmost column.}
	\begin{center}
		\begin{tabular}{lccc}
			\textbf{Parameter}& \textbf{Requirement}& \textbf{Goal}& \textbf{Proto-ChroMag}\\\hline
			Field of view& $2.5~\mathrm{R_\Sun}$& $2.5~\mathrm{R_\Sun}$& $2.5~\mathrm{R_\Sun}$\\
			Spatial resolution& $2\arcsec$& $1\arcsec$& $2.2\arcsec$\\
			Spectral coverage& \HeI\ $587.6~\nm$& Requirement and& Meets Goal\\
			& \HeI\ $1083.0~\nm$& \FeI\ $617.3~\nm$&\\
			&&\Halpha\ $656.3~\nm$&\\
			&&\CaII\ $854.2~\nm$&\\
			Spectral resolution& $23500$& $30000$& $32000$\\
			Cadence& $5~\minute$& $1~\minute$& $2~\minute$\\
			Polarimetric Accuracy& $10^{-3}$& $5\times10^{-4}$& $>5\times10^{-3}$\\\hline
		\end{tabular}
	\end{center}
	\label{tab:requirements}
\end{table}

ChroMag is an imaging polarimeter designed to measure on-disk chromosphere and off-disk prominence magnetic fields.
It is part of the planned COSMO suite of three instruments\footnote{See the COSMO homepage \url{http://www.cosmo.ucar.edu/} for details.}.
The centerpiece of COSMO is a $1.5~\meter$ refracting coronagraph that will measure coronal magnetic fields.
In order to exploit these measurements to their full potential for advancing science they should be considered in the context of the ambient plasma environment.
The COSMO large coronagraph is specifically dedicated to coronal magnetometry and hence requires supporting instruments to provide context measurements.
The COSMO white-light K-Coronagraph for coronal density measurements is currently under construction. 
It will be deployed to the NCAR Mauna Loa Solar Observatory (MLSO) facility in 2013.
ChroMag will also be deployed to the MLSO facility for regular operation in conjunction with the COSMO K-Coronagraph and the Coronal Multi-channel Polarimeter (CoMP) instruments, and later with the COSMO Large Coronagraph.

Besides having an important role in supporting the large Coronagraph, both the K-Coronagraph and ChroMag are highly valuable instruments in their own right.
ChroMag will provide important measurements of chromospheric magnetic fields at the base of the corona, as well as a measurement of magnetic field in filaments and prominences embedded in the corona.

The spectral line of choice for measurements of chromosphere and prominence magnetic field is the \HeI\ $1083.0~\nm$ triplet.
Similar to multi-line diagnostics of photospheric field, great benefit can be derived from combining polarimetry in the \HeI\ $1083.0~\nm$ triplet with the \HeI\ $D_3$ line at $587.6~\nm$ in prominences
	\cite{2009ApJ...703..114C}.
A key study for the design of ChroMag tested requirements on measurements by inverting synthesized lines and found that magnetic fields in prominences can be recovered with a filtergram instrument if the instrument has a narrow passband and good polarimetric accuracy
	\cite{CoSMOtech12}.

Several other lines are of interest.
The \Halpha\ line at $656.3~\nm$ is the diagnostic of choice for chromospheric structure and dynamics.
The well-known \CaII\ line at $854.2~\nm$ samples chromospheric magnetic field and structure in its core, while its wide wings provide a diagnostic of the atmosphere down to the photosphere.
Finally, a photospheric diagnostic is useful for cross-calibration and alignment with existing photospheric magnetometers.
The \FeI\ line at $617.3~\nm$ is a good choice as it is in use by the HMI instrument on board SDO.
\autoref{tab:requirements} summarizes the ChroMag instrument requirements and goals.

\section{DESIGN}
\label{sec:design}

The ChroMag instrument is currently in conceptual design phase
	\cite{CoSMOtech11}.
The baseline design consists of a refractive telescope with an aperture of $20~\cm$ that feeds a narrow-band Lyot filter imaging polarimeter.
The plan is that the future ChroMag instrument will be deployed to the NCAR MLSO facility for regular operation in conjunction with the COSMO K-Coronagraph and the CoMP instruments, and later with the COSMO Coronagraph.

A prototype instrument, Proto-ChroMag, is currently under construction.
The prototype instrument is planned primarily to demonstrate and verify the instrument concept and reduce risk in the instrument design.
Production of science-quality observations is a secondary objective.
Proto-ChroMag will have a smaller aperture than ChroMag proper, and hence will not meet the spatial resolution requirement.
Proto-ChroMag will also be a single-beam polarimeter, which means that it will be sensitive to atmospheric seeing conditions that induce crosstalk between intensity and polarization.
Since Proto-ChroMag will not be deployed to a high-quality site where good seeing conditions can be expected, the polarimetric accuracy is expected to suffer as a result.

\subsection{Optical Design}

\figuretwo{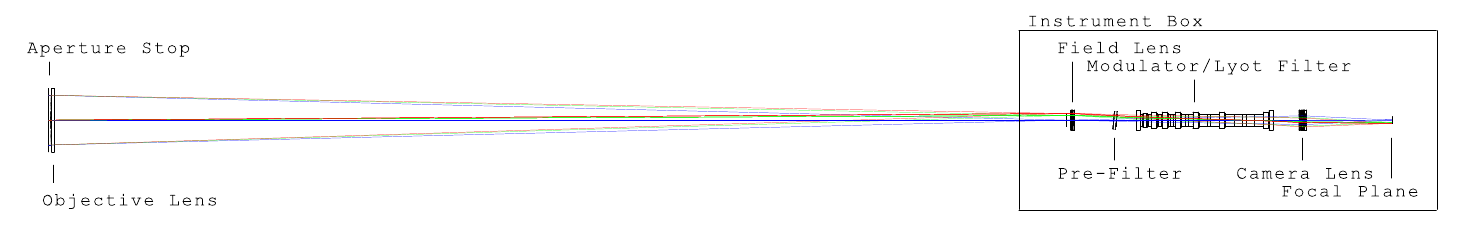}{fig:overview}{Overview of the Proto-ChroMag instrument optical design.}

The optical design of Proto-ChroMag is straight-forward.
An overview is shown in \autoref{fig:overview}.
The objective lens is taken from the decommissioned PICS\footnote{The Polarimeter for Inner Coronal Studies (PICS) instrument made observations of the disk and at the limb in \Halpha\ for studies of prominence and filament activity.  It was decommissioned in 2010.} instrument.
The field lens near the prime focus formats the beam so it can pass through the Lyot filter.
Five pre-filters for the lines given in \autoref{tab:requirements} are placed in a filterwheel just before the narrow-band filter.
The Lyot filter follows the design shown in \autoref{fig:fwhm_fsr}.
The four thinnest stages use calcite crystal taken from the birefringent \Halpha\ filter previously used in the decommissioned PICS instrument.
Thick pieces of optical-grade calcite are difficult to acquire.
The last two stages hence use two and four pieces of $22~\mm$ thick calcite, respectively.
Immediately behind the filter a camera lens creates an image with a plate scale of $1.1\arcsec$ per $6.5~\um$, appropriate for the baseline 5.5-megapixel Andor Neo sCMOS camera.
The polarimetric calibration optics (consisting of a polarizer and a waveplate in independent rotation stages) will be inserted between the objective lens and the field lens.
This means that the primary objective cannot be calibrated in the coarse of regular observations.
The telescope Mueller matrix will be measured in the lab by feeding it with light through a large-aperture calibration system previously used for the HMI instrument on SDO.

\subsection{Lyot Filter}

\figuretwo{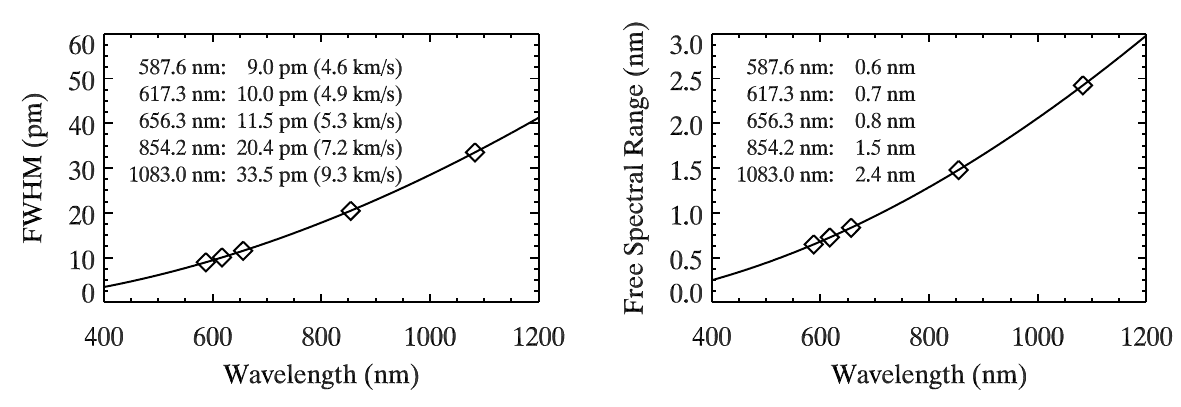}{fig:fwhm_fsr}{The full width at half maximum (left) and the free spectral range (right) of the Lyot filter designed for ChroMag.}

The heart of the ChroMag instrument is an electro-optically tunable wide-fielded narrow-band birefringent Lyot filter with a built-in polarimeter.
The filter is designed to meet the requirements in \autoref{tab:requirements}.
It will have six stages.
Each stage consists of an entrance polarizer, a calcite waveplate, an achromatic $\sfrac12$-wave plate, a second calcite waveplate, a liquid crystal variable retarder, and an exit polarizer.
The length of the calcite crystal is doubled in each successive stage, and the exit polarizer acts also as the entrance polarizer of the next stage.
The thinnest stage ($2\times1.375~\mm$ of calcite) determines the free spectral range of the filter.
The thickest stage has a total of $88~\mm$ of calcite crystal and determines the full width at half maximum of the transmission peak.
The theoretical performance of the filter design is shown in \autoref{fig:fwhm_fsr}.
The curves are not straight lines because of the dispersion of birefringence of calcite with wavelength.
As designed, this Lyot filter meets or exceeds the performance requirements given in \autoref{tab:requirements}.

\subsection{Modulator}

\begin{table}
	\caption{Specification of the ChroMag modulator.  Retardance is specified at $750~\nm$, and orientation is clockwise when looking at the detector with the zero point set by the orientation of the analyzing polarizer.}
	\begin{center}
		\begin{tabular}{lccc}
			\textbf{Device}& \textbf{Retardance}& \textbf{Orientation}\\\hline
			FeLC 1& $\sfrac12\lambda$& $0.00~\rad$\\
			FeLC 2& $\sfrac14\lambda$& $1.97~\rad$\\
			Retarder& $\sfrac14\lambda$& $1.92~\rad$\\
		\end{tabular}
	\end{center}
	\label{tab:modulator}
\end{table}

\figureone{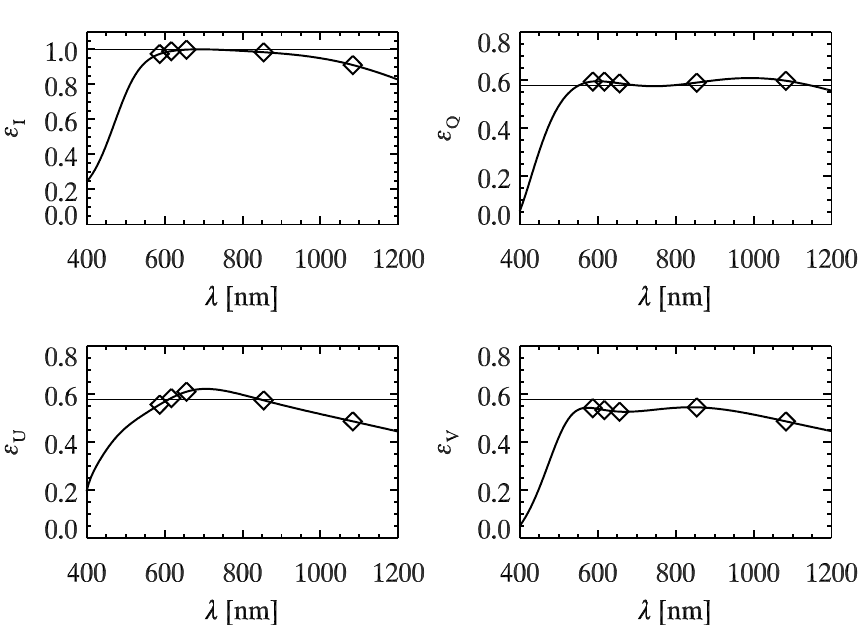}{fig:modulator}{The theoretical efficiency as a function of wavelength of the ChroMag polarimetric modulator.}

The built-in polychromatic modulator
	\cite{2010ApOpt..49.3580T}
consists of two Ferro-Electric Liquid Crystals (FeLCs) followed by a retarder (specified in \autoref{tab:modulator}).
The entrance polarizer of the first stage of the Lyot filter also acts as the polarization analyzer.
Since FeLCs and retarders are inherently chromatic, the modulator is not achromatic.
Its modulation matrix, in fact, is a strong function of wavelength.
It is however very nearly optimally efficient over the entire wavelength range, as is shown by the modulation efficiency curves in \autoref{fig:modulator}.

\section{FUTURE DEVELOPMENT}

ChroMag will differ from the prototype in several ways.
The main difference is that it will have a larger aperture, and the primary objective lens will be designed to be sufficiently achromatic over the $587.6$ to $1083.0~\nm$ range to function with an occulting disk at prime focus.
While ChroMag is not a coronagraph, an occulting disk will be inserted to block out most of the light from the solar disk when making measurements of off-limb structures such as prominences.
The calibration optics will be positioned in front of the telescope objective lens.
This permits the regular calibration of the full telescope Mueller matrix and will provide better polarimetric accuracy.

The Proto-ChroMag Lyot filter, currently under construction, is expected to also be used in the full instrument.
The ChroMag conceptual design calls for a dual-beam polarimeter that employs two identical Lyot filters
	\cite{CoSMOtech11},
which incurs significant additional cost over a single filter.
Several alternative options exist to mitigate the impact of seeing-induced cross-talk, such as incorporating a tip/tilt image stabilization system, or modifying the polarimetric modulator and filter so as to function as a dual-beam polarimeter using a single Lyot filter.
Proto-ChroMag will function as a pathfinder by allowing us to evaluate if dual-beam polarimetry is indeed necessary to meet the science requirements.

Proto-ChroMag will initially be deployed to the restored Boulder spar (to be installed at the NCAR Mesa Lab this summer).
It will be operated intermittently as a testbed instrument to verify key requirements and demonstrate the instrument concept.
The final ChroMag instrument will be deployed to the Mauna Loa Solar Observatory in Hawaii, where it will be used to observe the Sun daily in a synoptic fashion.
The data produced by ChroMag will be freely available to the community.

\acknowledgments

The National Center for Atmospheric Research is sponsored by the National Science Foundation. 


\end{document}